\documentclass[12pt,preprint]{aastex}

\usepackage{amsmath}
\usepackage{amssymb}
\usepackage{latexsym}

\usepackage{natbib}

\newcommand{\scinot}[2]{\ensuremath{#1 \times 10^{#2}}}

\newcommand{\paren}[1]{\left ( #1 \right )}

\newcommand{\curly}[1]{\left \{ #1 \right \} }

\newcommand{\parenfrac}[2]{\paren{\frac{#1}{#2}}}

\newcommand{\rwind}[1]{\ensuremath{r_{\text{wind}}^{#1}}}
\newcommand{\rhowind}[1]{\ensuremath{\rho_{\text{wind}}^{#1}}}
\newcommand{\vwind}[1]{\ensuremath{v_{\text{wind}}^{#1}}}
\newcommand{\Twind}[1]{\ensuremath{T_{\text{wind}}^{#1}}}

\newcommand{\lstop}[1]{\ensuremath{l_{\text{stop}}^{#1}}}

\newcommand{\theat}[1]{\ensuremath{t_{\text{heat}}^{#1}}}
\newcommand{\tcool}[1]{\ensuremath{t_{\text{cool}}^{#1}}}

\newcommand{\tanneal}[1]{\ensuremath{t_{\text{anneal}}^{#1}}}

\newcommand{\unit}[2]{\ensuremath{#1 \, #2}}

\newcommand{\cm}{\text{cm}}
\newcommand{\Msol}{\text{M}_{\odot}}
\newcommand{\cmpersecnp}{\text{cm} \, \text{s}^{-1}}
\newcommand{\grampercubiccmnp}{\text{g} \, \text{cm}^{-3}}
\newcommand{\kelvin}{\text{K}}
\newcommand{\AU}{\text{AU}}
\newcommand{\Mjup}{\text{M}_{\text{J}}}
\newcommand{\second}{\text{s}}
\newcommand{\micrometre}{\mu\text{m}}
\newcommand{\usk}{\,}
\newcommand{\hour}{\text{h}}
\newcommand{\erg}{\text{erg}}
\newcommand{\gramme}{\text{g}}
\newcommand{\reciprocal}[1]{{#1}^{-1}}

\begin{document}

\title{The Formation of Crystalline Dust in AGB Winds from Binary Induced Spiral Shocks}
\shorttitle{Crystalline Dust formation in AGB Winds}

\author{Richard~.G.~Edgar}
\affil{Department of Physics and Astronomy, University of Rochester,
Rochester, NY 14627}
\email{rge21@pas.rochester.edu}

\and

\author{Jason~Nordhaus}

\and

\author{Eric~G.~Blackman}

\and

\author{Adam~Frank}


\begin{abstract}
As stars evolve along the Asymptotic Giant Branch, strong winds are driven from the outer envelope.
These winds form a shell, which may ultimately become a planetary nebula.
Many planetary nebulae are highly asymmetric, hinting at the presence of a binary companion.
Some post-Asymptotic Giant Branch objects are surrounded by torii of crystalline dust, but there is no generally accepted mechanism for annealing the amorphous grains in the wind to crystals.
In this Letter, we show that the shaping of the wind by a binary companion is likely to lead to the formation of crystalline dust in the orbital plane of the binary.
\end{abstract}



\section{Introduction}

During the Asymptotic Giant Branch (AGB) phase, strong winds are driven from the outer envelope.
After $\sim 10^5$ yrs, the AGB envelope is expelled resulting in a proto-white dwarf surrounded by a circumstellar nebula.
As the hotter core of the star is unveiled, the nebula ionizes and becomes a planetary nebula (PN).
Most PNe are highly asymmetric, displaying complex morphological structures such as disks and bipolar jets \citep[e.g.][ and references therein]{2002ARA&A..40..439B}.
The engine driving the asymmetry is thought to begin during the AGB phase or shortly thereafter in the post-AGB phase.
A binary companion may be responsible for shaping the nebula.
Recent work suggests that most, if not all, PNe have incurred a binary interaction \citep{2004ASPC..313..515S,2005AIPC..804..169D,2006ApJ...645L..57S}.

The formation of equatorial torii, collimated bipolar jets and circumstellar disks may be a consequence of  additional energy and angular momentum supplied by the binary companion either through a common envelope phase or by directly shaping the AGB wind \citep{2006MNRAS.370.2004N,2007MNRAS.376..599N,2001ApJ...546..288B}.
A binary can also process the wide variety of dust species present in AGB winds \citep{2004ASPC..309..229W}.
Most of the grains are amorphous, with crystalline silicates also seen at 10-20\% abundance from stars with particularly high mass loss rates \citep{2001A&A...369..132K,2002MNRAS.332..513S}.
Some post-AGB systems possess a torus of crystalline dust at large radius \citep{2007BaltA..16..148G}.

Here, we discuss a method of producing the torus of crystalline dust: annealing in shocks induced by a binary companion.
A wind blowing past a binary companion will shock, as in Bondi--Hoyle--Lyttleton accretion \citep[see, e.g.][ for a description of Bondi--Hoyle--Lyttleton problem]{2004NewAR..48..843E}.
We will show that the shock temperatures can be sufficient to anneal the grains with the formed shock lying in the orbital plane of the binary.
Shock heating has been proposed before \citep{2003asdu.confE..63N}, but without a possible mechanism.

\section{Numerical Study}
\label{sec:numerics}

Our code is based on the \textsc{Flash} code of \citet{2000ApJS..131..273F}, an adaptive mesh refinement (AMR) code based around a Piecewise-Parabolic Method (PPM) hydrodynamics solver.\footnote{The source code is available at \url{http://flash.uchicago.edu/}}.
For this work, we use \textsc{Flash} in cartesian co-ordinates.
We have added a simple \textsc{nbody} solver, to model the binary.
The wind is modelled by resetting all grid cells within a distance \rwind{} of the primary to a density \rhowind{}, a temperature \Twind{}, and a radial velocity of \vwind{} with respect to the primary.
The gravitational effect of the bodies is subject to softening.
The softening length for the primary is less than \rwind{}; that of the secondary is chosen to be smaller than the expected Hoyle--Lyttleton radius \citep[see, e.g.][]{2004NewAR..48..843E}.
We do not include a jet \citep[cf][]{2004ApJ...600..992G}, since it is not relevant to the present study.

Close to each boundary there is a damping region, where the gas density and temperature are reduced to their ambient values.
We do this to ensure that waves cannot reflect from the boundaries.
We performed two sets of runs.
In the first, we used a \unit{1}{\Msol} primary, with $\rhowind{} = \unit{\ensuremath{10^{-14}}}{\grampercubiccmnp}$, $\Twind{}=\unit{\ensuremath{10^3}}{\kelvin}$, $\vwind{}=\unit{\scinot{3.8}{6}}{\cmpersecnp}$ and $\rwind{} = \unit{\scinot{2}{13}}{\cm}$.
The second set of runs contained a \unit{3}{\Msol} primary, with $\Twind{}=\unit{\scinot{2}{3}}{\kelvin}$, $\vwind{}=\unit{\scinot{6.5}{6}}{\cmpersecnp}$, and \rhowind{} and \rwind{} unchanged.
This choice of parameters ensures that the wind can escape from the system.
We refine the grid to ensure that $\rwind{}$ is always covered by 8 grid cells, while the secondary's Hoyle--Lyttleton radius is always covered by 4 grid cells.
The gas is assumed to be adiabatic, with $\gamma = 4/3$.

\section{Results}
\label{sec:results}

We shall start by discussing a sample calculation, with a \unit{1}{\Msol} primary, a \unit{0.25}{\Msol} secondary and an orbital semi-major axis of \unit{6}{\AU}.
This choice makes the Hoyle--Lyttleton radius of the secondary approximately \unit{\scinot{4.5}{12}}{\cm}.
Since we require at least four grid cells across the Hoyle--Lyttleton radius, our grid resolution is just over \unit{\ensuremath{10^{12}}}{\cm} in the vicinity of the planet's orbit.

\begin{figure}
\begin{center}
\plottwo{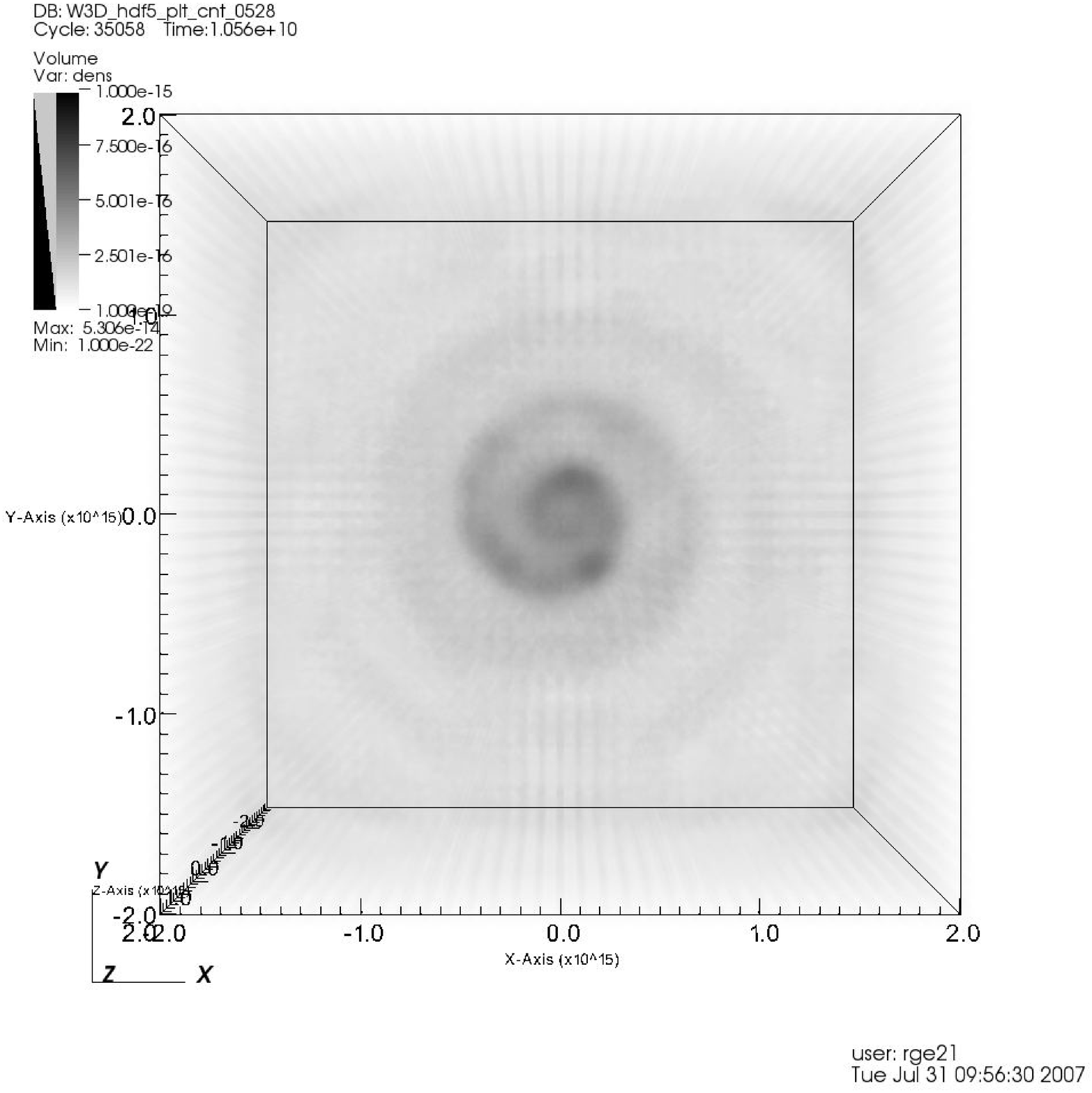}{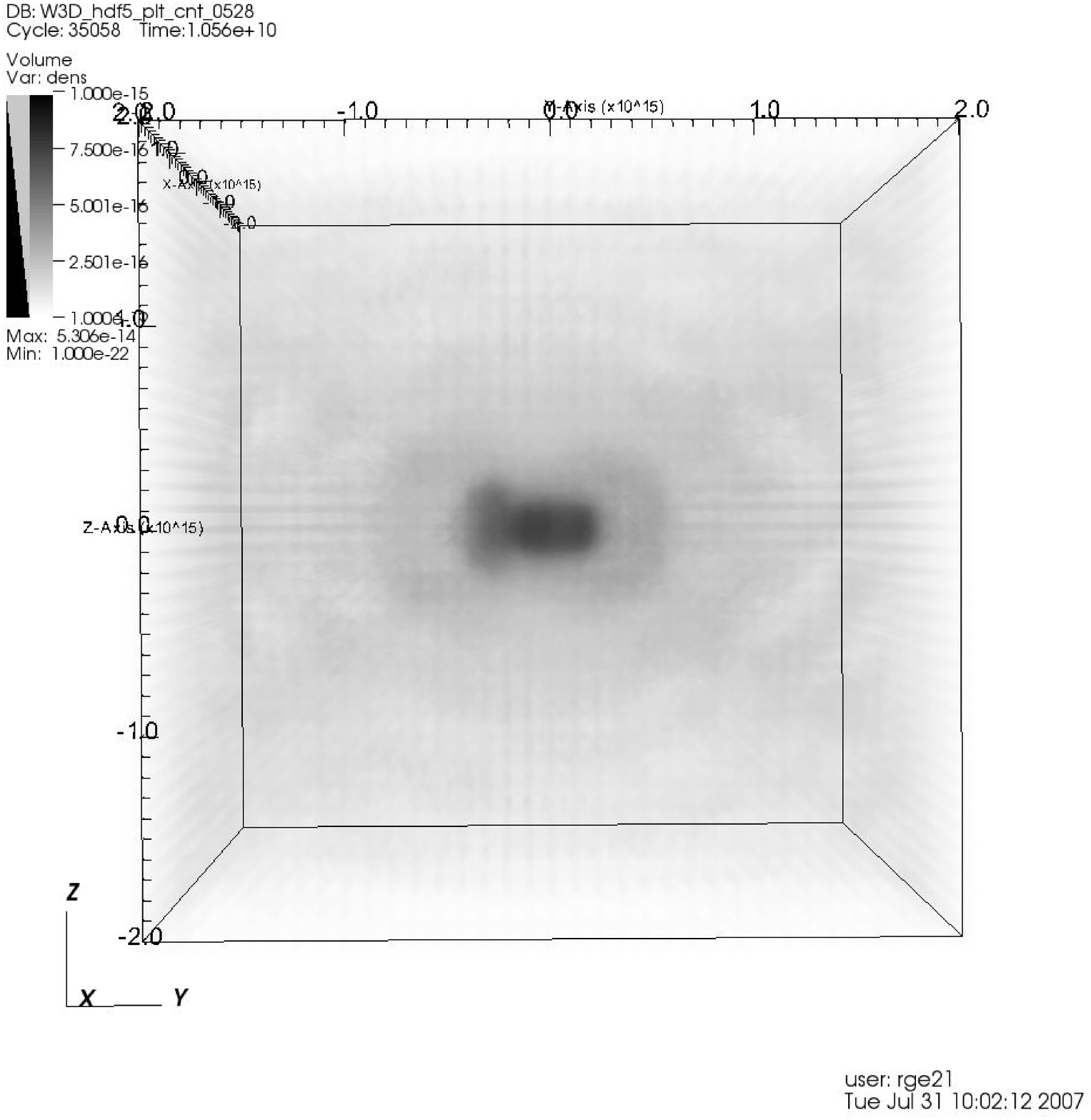}
\end{center}
\caption{Volumetric density renderings of the wind emitted by a \unit{1}{\Msol} primary, with a \unit{0.25}{\Msol} secondary in a \unit{6}{\AU} orbit. The view along the $z$ axis is shown on the left, that along the $x$ axis on the right.
The binary orbits in the $z=0$ plane}
\label{fig:densityq=25}
\end{figure}

\begin{figure}
\begin{center}
\plotone{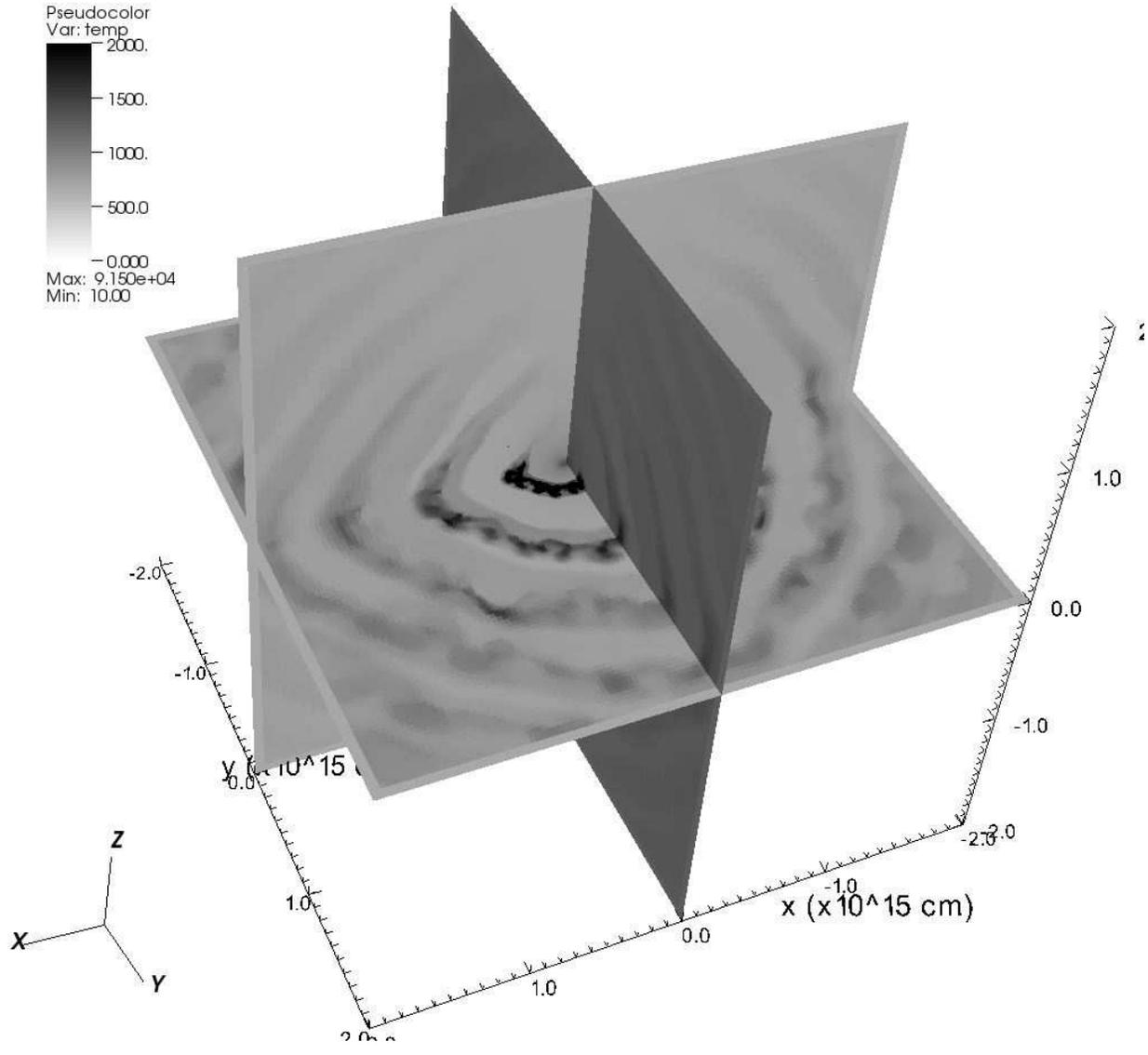}
\end{center}
\caption{Temperature structure of the wind  emitted by a \unit{1}{\Msol} primary, with a \unit{0.25}{\Msol} secondary in a \unit{6}{\AU} orbit}
\label{fig:temperatureq=25}
\end{figure}

In Figure~\ref{fig:densityq=25}, we show a volumetric density rendering of the system along two axes.
The spiral structure shown in the view along the $z$ axis is remarkably similar to that noted by \citet{2006A&A...452..257M} in \mbox{AFGL 3068}.
The tip of the spiral is attached to the secondary, forming a bow shock similar to that of Bondi--Hoyle--Lyttleton flow.
The shock is drawn into a spiral by the orbital motion.
The structure is also similar to that noted by \cite{1993MNRAS.265..946T} and \citet{1999ApJ...523..357M}.
Figure~\ref{fig:temperatureq=25} shows the temperature structure of the spiral.
Temperatures in excess of \unit{1000}{\kelvin} are generated in a narrow region $\sim \unit{\ensuremath{10^{13}}}{\cm}$ thick.
Although thin in comparison to the computational volume, this region is well resolved by our grid.

We ran a second numerical experiment, identical except for increasing the softening length of the secondary to a value substantially larger than the Hoyle--Lyttleton radius.
This isolated the effect of the secondary from the orbital motion of the binary.
The overall density structure was little changed, but the temperatures in the spiral dropped dramatically.
Therefore the orbital `recoil' of the primary is responsible for the macroscopic density structures, in support of the `piston' model of \citet{2007A&A...467.1081H}.\footnote{Compare especially their Figure~4 to our Figure~\ref{fig:densityq=25}}
However, the local deflection of the flow by the secondary drives the shock formation, and associated increase in temperature.

In a third numerical experiment, we increased the binary separation to \unit{10}{\AU}, but kept all the other parameters the same as the first.
The flow was similar to the first numerical experiment, with a slightly reduced peak temperature in the shock..

We also performed a fourth numerical experiment, identical to the first except for a \unit{100}{\Mjup} secondary (making a $q=0.095$ binary).
The spiral density structure was retained, but found that the peak temperature region was slightly lower, and spread over a smaller volume.

We performed a similar set of runs for a system with a \unit{3}{\Msol} primary, and found similar behaviour.
Since \vwind{} was increased to ensure that the wind could escape, the shock temperatures were higher.

\section{Discussion}
\label{sec:discuss}

Observations of post-AGB objects reveal crystalline dust to be concentrated in the midplane \citep{2007BaltA..16..145D}.
The presence of crystalline dust has been somewhat puzzling, since dust grains which form in AGB winds are expected to be amorphous.
Other authors have suggested that shock annealing might be responsible for crystallisation, but without exploring a mechanism for producing the shocks.
Meanwhile, many structures observed around post-AGB stars are attributed to binary evolution.
Our numerical experiments suggest that the binary and shock models should be combined: crystalline silicates form by annealing in the spiral shock formed as the primary's wind passes the secondary.
As the outward motion of the spiral shock slows, the arms will merge and appear as a torus.

\subsection{Grain Annealing}

There are a number of important scales which must be assessed to determine whether such annealing can occur.
First are the shock velocities.
Examining the output of our \textsc{Flash} runs,  we find that the velocities are always $\sim \unit{\ensuremath{10^6}}{\cmpersecnp}$.
Pre-shock, the velocities can be a factor of a couple higher.
Post-shock, the velocities can be a factor of a few lower.
In our calculations, the velocity is mainly radial at all times, so the crossing time of the high temperature region (\unit{\ensuremath{10^{12}}}{\cm} or so in radial extent in the midplane) is on the order of \unit{\ensuremath{10^6}}{\second}.

Next is the stopping distance of the grains by the gas.
Calculating the distance required for a grain to sweep up its own mass of gas, we find
\begin{equation}
\lstop{} = \parenfrac{4 \rho_{\text{dust}}}{3 \rho} a_{\text{dust}}
         = \scinot{1.3}{11}
           \parenfrac{\rho}{\unit{\ensuremath{10^{-15}}}{\grampercubiccmnp}}^{-1}
           \parenfrac{a_{\text{dust}}}{\unit{1}{\micrometre}} \usk \cm
\label{eq:lstop}
\end{equation}
where we have assumed a standard dust density of $\rho_{\text{dust}} = \unit{1}{\grampercubiccmnp}$.
For expected gas densities and dust sizes, this distance is sufficiently small to ensure that gas and dust are dynamically coupled; dust grains will not be blown out by radiation pressure independently of the gas.
The stopping timescale will be similar to the time required to heat the dust grains by collisions.
This is given by
\begin{equation}
\theat{} = \frac{\lstop{}}{v}
       = \scinot{1.3}{5}
         \parenfrac{\rho}{\unit{\ensuremath{10^{-15}}}{\grampercubiccmnp}}^{-1}
         \parenfrac{a_{\text{dust}}}{\unit{1}{\micrometre}}
         \parenfrac{v}{\unit{\ensuremath{10^6}}{\cmpersecnp}}^{-1}
         \usk \second
\label{eq:theat}
\end{equation}

\citet{2000ApJ...535..247H} studied the annealing of silicate grains as a function of temperature.
The process is very sensitive to temperature (it is controlled by the Boltzmann equation, with the rate $\propto \exp \curly{-E/k_B T}$), but if temperatures exceeding \unit{1067}{\kelvin} were reached, the amorphous grains would anneal to crystals on times $\tanneal{} \approx \unit{280}{\second}$.
At lower temperatures annealing stalled for around \unit{35}{\hour}, after an initial burst of crystallisation.
\citet{2002ApJ...565L.109H} applied this to spiral shocks in proto-planetary discs induced by gravitational instabilities, concluding that annealing would be possible.

The final timescale is that of cooling.
We have used a simple adiabatic equation of state for the gas.
In reality, the gas will be able to cool through a forest of molecular line transitions.
Computing these accurately is complicated, since many chemical reactions can occur.
To make an estimate, we use the cooling curves of \citet{2003A&A...404..267S}, who considered the wind of a pulsating star.
Using figure~4 of their work, we find
\begin{equation}
\tcool{} = \parenfrac{3 k_B T}{2 m_H \hat{Q}_{\text{rad}}}
         \approx \unit{\scinot{1.4}{5}}{\second}
\label{eq:tcool}
\end{equation}
where we have estimated the value of the cooling function for a temperature of \unit{1100}{\kelvin} and a density of \unit{\ensuremath{10^{-15}}}{\grampercubiccmnp}.
The cooling rate, $\hat{Q}_{\text{rad}}$ is roughly proportional to the density.

What are the implications for grain annealing?
First, $\tanneal{}$ is so short compared to the other timescales that, so long as temperatures in excess of \unit{1067}{\kelvin} are reached, we can assume that annealing is instantaneous.
If we could neglect post-shock cooling of the gas, then the crossing time for the whole shock is so long that annealing should occur even if the shock temperature is only \unit{1067}{\kelvin}.
However, cooling is likely to be significant, so successful annealing requires $\theat{} < \tcool{}$.
Specifically, we require
\begin{multline}
\frac{\theat{}}{\tcool{}} =
   0.9 \parenfrac{a_{\text{dust}}}{\unit{1}{\micrometre}}
       \parenfrac{\hat{Q}_{\text{rad}}}{\unit{\ensuremath{10^6}}{\erg\usk\reciprocal\gramme\reciprocal\second}} \times \\
       \parenfrac{\rho}{\unit{\ensuremath{10^{-15}}}{\grampercubiccmnp}}^{-1}
       \parenfrac{T}{\unit{1100}{\kelvin}}^{-1}
       \parenfrac{v}{\unit{\ensuremath{10^6}}{\cmpersecnp}}^{-1}
    < 1
\end{multline}
Note that $\hat{Q}_{\text{rad}}$ is approximately proportional to density, so this ratio is fairly insensitive to the gas density.
Smaller grains will heat (and hence anneal) faster.
We calculated \theat{} for \unit{1}{\micrometre} grains, but this is an upper limit.
We expect the actual grains to be smaller, perhaps as small as \unit{0.1}{\micrometre}, making the heating time an order of magnitude less.
Such grains would anneal easily - particularly when one recalls that each grain trajectory will encounter the spiral shock multiple times.

\subsection{Shock Temperature Scaling}

We now estimate how the shock temperature scales with the system parameters.
The problem may be split into two parts:
the wind from the primary may be modelled as a spherical Bondi wind \citep[see, e.g. chapter~2 of][]{2002apa..book.....F}.
A shock is then induced in this wind by the presence of the secondary.
Post-shock temperatures of \unit{1067}{\kelvin} will anneal the dust grains.

A most important parameter in a Bondi wind is the sonic point.
This occurs at a radius
\begin{equation}
r_s = \frac{GM}{2 c_s^2(r_s)} \approx \scinot{7.5}{13} \parenfrac{T(r_s)}{\unit{\ensuremath{10^4}}{\kelvin}}^{-1} \parenfrac{M}{\unit{1}{\Msol}} \usk \cm
\end{equation}
In our winds $T \sim \unit{1000}{\kelvin}$ or less, implying that the sonic point will be comfortably outside the orbit of the secondary.
Since we construct our initial conditions to be escaping and supersonic, it is reasonable to assume that our wind solution should always be supersonic (the Type~4 solutions of \citeauthor{2002apa..book.....F}).

The Bondi solution must be obtained numerically.
To proceed, we shall assume that the Mach number of the flow remains constant and that the wind velocity remains close to the escape velocity.
Both assumptions are correct for highly supersonic Bondi flows.
The launch Mach number of the wind is then
\begin{equation}
\mathcal{M}_{\text{launch}} \approx 11.5
        \parenfrac{M}{\unit{1}{\Msol}}^{\frac{1}{2}}
        \parenfrac{\rwind{}}{\unit{\scinot{2}{13}}{\cm}}^{-\frac{1}{2}}
        \parenfrac{\Twind{}}{\unit{1000}{\kelvin}}^{-\frac{1}{2}}
\label{eq:MachLaunch}
\end{equation}
and
\begin{equation}
T(r) \approx \Twind{} \parenfrac{M}{\unit{1}{\Msol}}
                      \parenfrac{r}{\unit{\scinot{2}{13}}{\cm}}^{-1}
\label{eq:Tprofile}
\end{equation}
The flow then shocks due to the presence of the secondary.
We can use the Rankine-Hugoniot equations to find
\begin{equation}
\frac{T_2}{T_1} = \frac{ \paren{2 \gamma \mathcal{M}^2 - \paren{\gamma-1}}  \paren{ \paren{\gamma-1}\mathcal{M}^2 + 2 } } {\paren{\gamma+1}^2 \mathcal{M}^2}
\label{eq:Tshock}
\end{equation}
where $T_1$ and $T_2$ are the pre- and post-shock temperatures respectively, and $\mathcal{M}$ is the preshock Mach number.
For a strong shock $\mathcal{M} \gg 1$ in a $\gamma=4/3$ gas, Equation~\ref{eq:Tshock} becomes $T_2 \approx 0.16 \mathcal{M}^2 T_1$.

The Mach number of the shock is \emph{not} $\mathcal{M}_{\text{launch}}$, since we must add in the orbital motion of the companion in quadrature to obtain the total relative velocity of the gas.
This is comparatively straightforward, since the velocity of a body in a circular orbit is a factor $\sqrt{2}$ smaller than the escape velocity.
This implies that $\mathcal{M}^2 \approx 1.5 \mathcal{M}_{\text{launch}}^2$.
Combining Equations~\ref{eq:MachLaunch}, \ref{eq:Tprofile} and~\ref{eq:Tshock}, we see that
\begin{equation}
T_2 \approx 1200 \parenfrac{M}{\unit{1}{\Msol}}^2
          \parenfrac{\rwind{}}{\unit{\scinot{2}{13}}{\cm}}^{-1}
          \parenfrac{r_{\text{orb}}}{\unit{\ensuremath{10^{14}}}{\cm}}^{-1}
\label{eq:TshockScale}
\end{equation}
This assumes that the wind velocity is always equal to the escape velocity.
Our numerical experiments had \vwind{} slightly larger than the escape velocity, leaving some `excess' velocity which Equation~\ref{eq:TshockScale} does not take into account.
Consequently, Equation~\ref{eq:TshockScale} is rather more sensitive to $r_{\text{orb}}$ than in our numerical experiments.
We emphasise that this calculation is only a rough estimate of the temperatures reached, since we saw a wide range of temperatures in our numerical experiments.
At the tip of the bow shock the temperature was much higher (this material would probably be accreted by the secondary anyway).
Moving along the spiral arms, we found that the temperatures rapidly dropped to values similar to those predicted by Equation~\ref{eq:TshockScale}.

Our numerical simulations indicate that the shock temperatures depend weakly on the secondary mass -- two orders of magnitude in secondary mass lead to $\sim 20\%$ change in temperature.
This is not included in Equation~\ref{eq:TshockScale} above, but the simulations are however consistent with the much stronger predicted dependence on the primary mass.
There are also two important  scales for the secondary in addition to its own radius, namely the Hoyle-Lyttelon radius and the Roche radius.
In our simulations and in the analytic scaling above we implicitly assume that the  Hoyle--Lyttleton radius is smaller than the Roche radius.
In cases where the reverse applies, we would expect an accretion disk to form close to the secondary.
The shock structure and temperature may depend somewhat on the ratio of these two radii, which we have not pursued in the present work.

What limits are appropriate for $T_2$?
As we have already mentioned, we require $T_2 > \unit{1067}{\kelvin}$ for annealing to occur.
However, it must not be too high, or the grains will vapourise - \unit{2000}{\kelvin} is a good upper limit.
Equation~\ref{eq:TshockScale} then provides rough constrains on the systems which can produce crystalline dust torii.
For a \unit{1}{\Msol} star, we would require $\scinot{0.6}{14} < r_{\text{orb}} < \unit{\scinot{1.1}{14}}{\cm}$ (assuming $\rwind{} = \unit{\scinot{2}{13}}{\cm}$).
If the primary were a \unit{3}{\Msol} star, the limits become $\scinot{5.4}{14} < r_{\text{orb}} < \unit{\scinot{1}{15}}{\cm}$ (assuming the same $\rwind{}$ value).

\section{Conclusion}
\label{sec:conclude}

We have demonstrated that a binary companion to an AGB star can create a torus of crystalline dust.
The crystalline dust is formed by the annealing of amorphous grains in the spiral shock induced by the companion.
Such a torus is likely to be slowly expanding, and not in Keplerian orbit around the system.

Although we are certain that our basic mechanism is robust, we have only made rough estimates of cooling and heating.
We have demonstrated that the relevant timescales should permit annealing, but more work is needed.
Future calculations should incorporate gas cooling.
This is not straightforward, since the relevant temperatures and timescales imply non-equilibrium chemistry.
Ideally, the dust should also be incorporated into the code as a separate, coupled component.

Crystalline dust torii provide strong evidence for binary interactions in AGB winds.


\acknowledgements

We acknowledge support from HST grant AR-10972, NSF grants AST-0406799, AST-0406823, and NASA grants ATP04-0000-0016 (NNG05GH61G) and NNG04GM12G.
This work is supported in part by the U.S. Department of Energy under Grant No. B523820 to the Center for Astrophysical Thermonuclear Flashes at the University of Chicago.
The computations presented here were performed on the \texttt{lonestar} cluster of the Texas Advanced Computing Center, using time granted through TeraGrid under project \mbox{TG-AST070018T}.
We are grateful to Garrelt Mellema, for helpful comments about gas cooling




\clearpage

\clearpage

\end{document}